\documentclass[12pt]{elsart}
\usepackage{epsfig}\def\epsfg#1#2{\epsfig{file=#1.ps,width=#2}}
\def\boldsymbol#1{\mbox{\boldmath ${#1}$}}

\begin{document}

\begin{frontmatter}

\title{The constituent quark as a soliton in a linear $\sigma$-model}

\author[PeF,IJS,EM1]{B. Golli} and
\author[FMF,IJS,EM2]{M. Rosina}

\address[PeF]{Faculty of Education, University of Ljubljana, 
              Ljubljana, Slovenia}

\address[FMF]{Faculty of Mathematics and Physics, 
              University of Ljubljana, Ljubljana, Slovenia}

\address[IJS]{J.~Stefan Institute, Jamova 39, P.O.~Box 3000,
              1001 Ljubljana, Slovenia}

\thanks[EM1]{E-mail: bojan.golli@ijs.si}
\thanks[EM2]{E-mail: mitja.rosina@ijs.si}

\date{22 October 1996}

\begin{abstract}
We study a stable, soliton-like solution in the linear 
$\sigma$-model where the chiral fields are coupled to a 
single massless nonstrange quark.
We investigate a possibility that this solution represents
the constituent quark of Georgi and Manohar.
We show that its properties 
are indeed consistent with nucleon observables.
Furthermore, we derive chiral meson exchange potentials 
between such objects which are similar to recently used
potentials in the constituent quark models.
\end{abstract}

\end{frontmatter}


\section{Introduction}

The successes of the nonrelativistic constituent quark models 
stimulate the study of the microscopic structure of 
the constituent quark.
Such a structure can not only influence predictions for many 
observables of the nucleon but also offer a possibility 
to derive an important part of the effective interaction 
between constituent quarks.

The idea that the constituent quark is represented by a
current quark surrounded by a chiral field, stems from Georgi 
and Manohar \cite{MG} and has been further elaborated 
by Cheng and Li \cite{ChengLi}, and by Baumgartner, Pirner, 
K\"onigsmann and Povh \cite{Pirner} in order to account for 
the parton counting in deep inelastic scattering (related to 
the flavour asymmetry of sea quarks as given by the Gottfried 
sum; and to strangeness and spin content of the nucleon).
We offer an explicit candidate for such a constituent quark: 
a soliton-like solution of a coherent state of pions and 
$\sigma$-mesons around a massless quark in
the linear $\sigma$-model.

The choice of a mesonic rather than gluonic description of 
the structure of the constituent quark is based on the experience 
that the scale for chiral symmetry breaking and emergence of 
chiral Goldstone bosons appears at lower energies than the 
confinement scale where gluons dominate.
The proposed picture may be a good representation of constituent 
quarks in the way they contribute to static properties of hadrons 
and to the low lying energy spectrum. 
It has been recently substantiated by several groups 
\cite{Riska,Paco,Fdelar,Glozman,Yoshi} which demonstrated 
that the meson exchange effective potential between constituent 
quarks yields much more reasonable predictions for the baryon 
spectroscopy and the baryon-baryon interaction than the gluon 
exchange potential.

\section{The model}

One of the simplest models describing the spontaneous breaking 
of the chiral symmetry is the linear $\sigma$-model. 
In the non-strange sector it involves u and d quarks, a
triplet of pions and the $\sigma$-meson.
The lagrangian density of the model \cite{Sigma,BB,KRS} takes 
the form:
\begin{equation}
    \mathcal{L} = {\rm i}\bar{\psi}\gamma^\mu \partial_\mu\psi
             + g\bar{\psi}
   (\hat{\sigma}+{\rm i}\pol{\tau}\cdot\hat{\pol{\pi}}\gamma_5)\psi
 + \half\partial_\mu\hat{\sigma}\partial^\mu\hat{\sigma}
 + \half\partial_\mu\hat{\pol{\pi}}\cdot\partial^\mu\hat{\pol{\pi}} 
 - \mathcal{U}(\hat{\sigma},\hat{\pol{\pi}}),
\label{lagrangian}
\end{equation}
where different contributions in order represent the free quark 
part, the linear quark-meson interaction, the free meson part
involving the $\sigma$-mesons and the pions, and the meson 
self-interaction:
\begin{equation}
    \mathcal{U} = {m_\sigma^2 - m_\pi^2\over 8 f_\pi^2} 
    \left(\hat{\sigma}^2 + \hat{\pol{\pi}}^2 - f_\pi^2\;
      {m_\sigma^2 - 3m_\pi^2\over m_\sigma^2 - m_\pi^2}\right)^2
     - f_\pi m_\pi^2\hat{\sigma}\;.
\label{potential}
\end{equation}
The final term in (\ref{potential}) is introduced to explicitly 
break the symmetry giving a finite mass to the pion.
The model contains only one adjustable parameter, $g$.
The other three parameters, the chiral meson masses 
and the pion decay constant, are fixed to the corresponding
experimental values; we use $m_\pi=139$~MeV and $f_\pi=93$~MeV,
for the mass of the $\sigma$-meson we have taken a typical
value, the recently suggested $m_\sigma=860$~MeV \cite{expsigma}.
The results, however, only weakly depend on the $\sigma$-meson mass.

We use coherent states to describe the pion and the $\sigma$-meson 
clouds, and, furthermore, the ``hedgehog'' ansatz  for the pion 
field and the current quark state
-- in complete analogy to the approaches in which
the nucleon is described as a meson cloud surrounding the three
valence quarks  \cite{BB,KRS,GR,Mike}.
The intrinsic state takes the form
\begin{eqnarray}
 |H\rangle &=& \mathcal{N}
{\rm exp}\left\{\sum_m (-1)^{1-m}\int \d k
     \sqrt{2\pi\omega_k/3}\, k\,\Pi(k) a^\dagger_{1m,-m}(k)\right\}
\nonumber\\
   &&\times {\rm exp}\left\{\int \d k\sqrt{2\pi\widetilde{\omega}_k}
          \,k\,\Sigma(k) \widetilde{a}^\dagger(k)\right\} 
    {1\over\sqrt{2}}\left(|u\downarrow\rangle - |d\uparrow\rangle
                    \right)\;.
\label{hedgehog}
\end{eqnarray}
Here $\widetilde{a}^\dagger(k)$ and $a^\dagger_{lmt}(k)$ are the 
creation operators for the $\sigma$-meson and the pion, respectively, 
in the spherical basis and $t$ is the third component of isospin;
$\omega_k^2=k^2+m_\pi^2$, $\widetilde{\omega}_k^2=k^2+m_\sigma^2$,
and $u\downarrow$ and $d\uparrow$ represent the u (d) quark 
with spin down (up), respectively.
Only $s$-wave $\sigma$-mesons and $p$-wave pions are coupled to 
the quark. The functions $\Sigma(k)$ and $\Pi(k)$ are related 
to the expectation values of the field operators as 
$\langle H|\hat{\sigma}(\vec{r})|H\rangle=\sigma(r)$ and
$\langle H|\hat{\pi}_t(\vec{r})|H\rangle=\pi(r)\hat{\vec{r}}_{-t}$,
where $\sigma(r)-f_\pi\equiv\tilde{\sigma}(r)$ and $\pi(r)$ 
are the Fourier transforms of $\Sigma(k)$ and $\Pi(k)$, respectively. 
The physical states are obtained by performing the Peierls-Yoccoz 
projection
\begin{equation}
 |J,M_J,T,M_T\rangle = P^{\mathrm{spin}}_{JM_J}
                       P^{\mathrm{isospin}}_{TM_T}|H\rangle \;.
\label{projection}
\end{equation}
Using the grand spin symmetry of the hedgehog, the evaluation
of physical quantities between the projected states is
considerably simplified; details of the calculation technique
as well as a discussion of the validity of the hedgehog
approximation can be found in \cite{CFGR}. 

The {\it ansatz\/} (\ref{hedgehog},\ref{projection}) leads to 
a system of coupled nonlinear differential equations which are 
solved using the package {\it COLSYS\/} \cite{colsys}.
We also impose a constraint that ensures the correct asymptotic
behaviour of the pion field which is, otherwise, not automatically 
satisfied in this type of variational approach \cite{AFGR}. 
We have found a selfconsistent solution with $T=J={1\over2}$
for the range $5.7<g<13.3$ (Fig.~1); we shall call
this solution the {\it quark soliton\/} (QS)%
\footnote{The soliton in the $\sigma$-model with $N_c=1$ was 
first discussed by V.~Soni in the context of the maximum 
fermion mass \cite{Vikram}.}.
For lower values of $g$, the trivial solution $\sigma=f_\pi$, 
$\pol{\pi}=0$ has a lower energy while for larger $g$, the valence 
orbital already drops into the Dirac sea.

\begin{figure}
\begin{center}
\epsfg{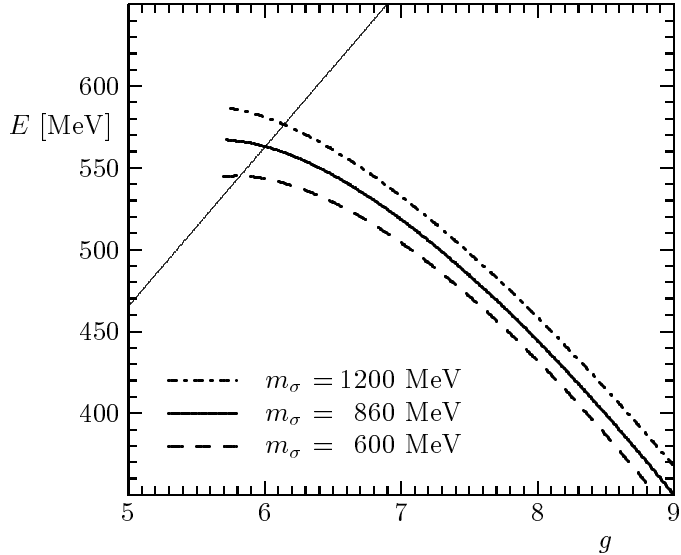}{90mm}
\end{center}
\caption{The energy of the quark soliton as a function of the coupling 
constant $g$ for three values of the $\sigma$-meson mass.
The straight line represents the energy of the trivial solution,
$E_0=gf_\pi$.}
\end{figure}

In the constituent quark models the quark mass is usually taken 
between 310~MeV and 390~MeV; from Fig.~1 one might naively expect 
to reproduce these values for $g\sim 9$.
However, the model possesses another nontrivial approximate solution 
representing the nucleon -- the {\it nucleon soliton\/} (NS) -- 
in which the three current quarks are surrounded by a common pion 
and $\sigma$-meson cloud \cite{GR,Mike} and here the experimental 
nucleon mass can be reproduced for $g$ below $\sim 6$.
Since the soliton energy decreases with $g$ higher values 
are not physically admissible for the QS either.
For $g\sim 6$ the energy of the soliton is rather independent 
of $g$ and is around 550~MeV which is considerably higher than 
a typical mass of the constituent quark.
This may not be a too serious shortcoming of the model for the
following two reasons:
(i) as we shall show in the following, there exist a strong 
attractive interaction between the QSs due to meson exchange 
which considerably lowers the energy of the three-QS system,
(ii) the present solution is strongly localized and 
the energy of the soliton contains a large contribution of
the kinetic energy of the spurious center-of-mass motion.
Both effects allow the existence of a solitonic solution 
also for $g$ below 5.7;
we may even expect that the physically sensible solution
should be sought below 5.

In order to investigate point (ii) and to be able to obtain 
a stable one-quark soliton solution in the physically relevant
range of $g$ it would be important to perform, in addition to
the angular momentum, also the linear momentum projection.
Techniques for linear momentum projection
were developed in soliton models \cite{Wilets} and for 
combined linear and angular momentum projection \cite{Neuber}.
The resulting lowering of the soliton energy is rather strong,
in the case of the QS even exaggerated.
We have found out that the strong effect is primarily due to 
the assumption that the negative energy Dirac sea orbitals
are plane waves, while in fact they are distorted by the soliton.
This distortion diminishes the effect of projection, especially
since some negative energy states become localized.
In order to get reliable center-of-mass corrections, Dirac sea 
polarization should be implemented.
This work is in progress.

The main challenge of the model is to build hadrons from
the quark solitons.
In the following sections we shall present a simple model 
of the nucleon assuming that the QSs preserve their 
identity in the baryon.
This is a very strong assumption 
since it implies 
that the system of three interacting QSs
is energetically more favourable than the NS.
It would have to be investigated using more sophisticated approaches 
(e.g. removal of center-of-mass motion of a single QS, solving 
the Faddeev equations for the three quark system, $\ldots$)
which are far beyond the scope of this letter.
%Though the success of the constituent quark model strongly 
%supports the picture of almost independent constituent quarks 
%it is worth investigating whether such a picture follows from 
%our simple model.

Though we are not able in the present approach to generate
stable solitons in the physically interesting range, 
it is still interesting to
study their properties at somehow larger $g$.
As we shall show in the following, several properties depend 
only weakly on the parameter $g$ and we may expect that they 
will, at least qualitatively, persist when more sophisticated
approaches are used.

\section{Static properties}

In Table 1 we summarize the static properties of the QS.
The average number of pions ($n_\pi$) is not a physical 
observable, nonetheless, it is an important tool in analyzing 
the flavour and spin content of the nucleon.
The violation of the Gottfried sum rule, the Drell-Yan 
process and the deep inelastic lepton-nucleon scattering 
measuring various quark contributions to the nucleon spin 
can well be understood assuming the quarks are surrounded 
by a pion cloud \cite{ChengLi,Pirner}.
The parameter of this analysis, $a$, representing the 
probability for the $\pi^+$ pion in the cloud, would lead to 
$n_\pi=1.5a$ if two- and more pion configurations are neglected.
Since the determination of this parameter is model dependent
and still very inconclusive, our values in Table 1 are
consistent with $a=0.1$ from Ref.~\cite{ChengLi} and $a=0.246$ 
from Ref.~\cite{Pirner}.

\begin{table}[hbt]
\vspace*{6pt}
\caption{Static properties of the constituent quark. 
For comparison, the nucleon properties deduced from 
the three-QS system ($3\times\hbox{QS}$), 
those calculated for the nucleon soliton (NS), 
and the experimental values are given.
The notation $t={1\over2}$ ($t=-{1\over2}$) refer to U (D) 
constituent quark or to proton (neutron), respectively.}
\begin{center}
\vspace*{6pt}
\begin{tabular}{crrrrrr}
\hline\hline
Quantity &\multicolumn{3}{c}{Quark Soliton}
         &\quad$3\!\times\!\hbox{QS}$  & NS & Exp. \\
         &$g=5.72$ & $g=6$ & $g=7$ & $g=6$ & $g=6$ & (Nucl.)\\
\hline
$n_\pi$     & 0.17 & 0.23 & 0.34 & 0.69 & 1.10 & --  \\
$g_{\pi QQ}$ ($g_{\pi NN}$)   
            & 7.58 & 7.72 & 7.96 & 12.9 & 17.0 & 13.5 \\
$g_A$       & 0.95 & 0.96 & 0.97 & 1.60 & 1.79 & 1.26 \\
$\mu_{t= {1\over2}}$ 
  [n.m.]& 0.99  & 1.03  & 1.07  & 1.61  & 2.82  &2.79 \\
$\mu_{t=-{1\over2}}$ 
  [n.m.]&$-0.67$&$-0.73$&$-0.82$&$-1.32$&$-2.49$&$-1.91$\\
$r^2_{t= {1\over2}}$ 
  [fm$^2$]& 0.30  & 0.25  & 0.23  &0.48   &0.54 &0.72\\
$r^2_{t=-{1\over2}}$ 
  [fm$^2$]&$-0.17$&$-0.14$&$-0.13$&$-0.04$&$-0.11$&$-0.11$\\
\hline\hline
\end{tabular}
\end{center}
\end{table}

The value for the constituent quark-pion coupling constant 
$g_{\pi QQ}$ leads to a good agreement of  
$g_{\pi NN}\;(={5\over3}g_{\pi QQ})$ with the experimental value.
The contribution of the meson self-interaction 
is significant ($\approx 25$~\%).

The axial current acquires an important contribution from 
the $\sigma$ and the pion field, and using a simple relation
$g_A(\mathrm{nucleon})={5\over3} g_A(\mathrm{quark})$ leads 
to too large a value of $g_A$ for the nucleon.
The meson fields of constituent quarks may considerably alter
in the nucleon giving rise to effective exchange (axial) 
currents which may eventually reduce this value. 

The magnetic moments are too small, yet not too far from the 
corresponding Dirac values (e.g. 
$\mu_U^{\mathrm{Dirac}}/\mu_N={2\over3}M_p/M_Q=1.12$).
We can therefore expect that reducing the mass of the QS in
an improved computational scheme would increase the proton 
and the neutron magnetic moments, 
$\mu_{p(n)}={1\over2}(\mu_U+\mu_D)\pm{5\over6}(\mu_U-\mu_D)$,
to more acceptable values.
Similarly as for $g_A$, the effective exchange currents may 
yield an important contribution also in this case.

The calculated charge radii pertain to the size of the 
constituent quark.
To obtain the nucleon charge radius we 
have taken into account the orbital motion of the constituent
quarks in the nucleon as discussed in the next section.
The nucleon charge radius 
would be considerably increased if the vector mesons
were introduced; our values are therefore not inconsistent
with the experiment.

\section{Effective potentials between constituent quarks}

We have calculated the chiral meson exchange effective potentials 
between constituent quarks using the Born Oppenheimer approximation.
While the assumption that the QS behaves as a static source of 
mesons can be justified in the case of the pion because of its 
small mass compared to that of the QS, 
it is certainly not appropriate for the $\sigma$-meson.
Lacking a more consistent approach in which recoil effects are
taken into account, the resulting $\sigma$-exchange potential 
should be regarded only as a crude approximation.

We give here explicit expressions for the contributions 
arising from the quark-meson interaction in (\ref{lagrangian}) 
which dominate.
The meson self-interaction (\ref{potential}) yields the pion 
and $\sigma$-exchange potentials of similar shapes as the 
corresponding ones arising from the quark-meson interaction;
they merely increase the strength of the total potentials 
by some 20~\%.

The quark contribution to the static $\sigma$-exchange potential 
is simply:
\begin{eqnarray}
  V^q_\sigma(r)& = & g\int \d^3\vec{r}' 
    \bar{\psi}_a(\vec{r}')\psi_a(\vec{r}')
    \tilde{\sigma}_b(|\vec{r}-\vec{r}'|)
\nonumber\\
    &=& g\int_0^\infty \d r'\,{r'}^2(u(r')^2-v(r')^2)
   \half\int_{-1}^1\d \cos\vartheta\, 
   \tilde{\sigma}(|\vec{r}-\vec{r}'|)\;,
\label{sigmaex}
\end{eqnarray}
where the index $a$ refers to the QS positioned at the origin
and $b$ to the QS positioned at $\vec{r}$, $u$ ($v$)
is the quark bispinor upper (lower) component, and $\vartheta$ 
is the angle between $\vec{r}$ and $\vec{r}'$.
The potential, including also the contribution 
from (\ref{potential}), is attractive for all $r$. 
Its value at the origin is $V_\sigma(0)=-320$~MeV and has 
a harmonic shape with $k=1.2$~GeV/fm$^2$ at small $r$.
Asymptotically it has a typical Yukawa behaviour with the
mass $2m_\pi$ (rather than $m_\sigma$) as a consequence of
remaining close to the chiral circle.

In order to write down the pion exchange potential we first
notice that the pion field of the QS can be written 
in the form:
\begin{equation}
  \pol{\pi}_b(\vec{r}) = {1\over3}\pi(r)\,\hat{\vec{r}}\cdot
    \vec{\Sigma}_b \,\pol{T}_b\;,
\label{pionfield}
\end{equation}
where $\vec{\Sigma}$ and $\pol{T}$ are, respectively, the spin 
and isospin operators acting on dressed (constituent) quarks.
Similarly, the quark-pion interaction term in (\ref{lagrangian}) 
can be manipulated:
\begin{equation}
  {\rm i}\bar{\psi}_a\pol{\tau}_a\gamma_5\psi_a 
\to
 2u(r)v(r)\, \hat{\vec{r}}\cdot\vec{\sigma}_a \pol{\tau}_a 
\to
 2u(r)v(r)\,
 {\langle Q||\sigma\tau||Q\rangle \over 
   \langle Q||\Sigma T||Q\rangle}\,
  \hat{\vec{r}}\cdot\vec{\Sigma}_a \pol{T}_a
\label{quarksource}
\end{equation}
yielding the quark contribution to the pion exchange potential:
\begin{equation}
  V^q_\pi(r) =  {2g\over 3} \langle Q|\sigma_0 \tau_0|Q\rangle
 \int \d^3\vec{r}'u(r')v(r')\, \pi(\rho)\,
 \hat{\vec{r}}'\cdot\vec{\Sigma}_a\,
 \hat{\vec{\rho}}\cdot\vec{\Sigma}_b\, 
  \pol{T}_a\pol{T}_b
\label{pionex}
\end{equation}
where $\vec{\rho}=\vec{r}-\vec{r}'$, and $\vec{\sigma}$ 
and $\pol{\tau}$ now act on current quarks.
The potential (\ref{pionex}) contains the scalar as well 
the tensor part.
The scalar part of the pion exchange potential including 
also the contribution from the meson self-interaction 
(\ref{potential}) can be written in the form
$V_\pi(r)\,\vec{\Sigma}_a\!\!\cdot\!\!\vec{\Sigma}_b\,
\pol{T}_a\pol{T}_b$; $V_\pi(r)$ is shown in Fig.~2 and 
compared to some commonly used potentials in the constituent 
quark model calculations of the baryon spectra 
\cite{Paco,Fdelar,Glozman}.
We reproduce well its asymptotic behaviour -- which reflects 
the fact that the pion coupling constant is also well reproduced 
in the model -- while the internal part has a too large range.
It could be reduced by going to stronger $g$ but this 
is physically not desirable.
The large range is partially due to spurious center-of-mass 
motion so we may expect that its size would be reduced 
by removing this effect.

\begin{figure}
\begin{center}
\epsfg{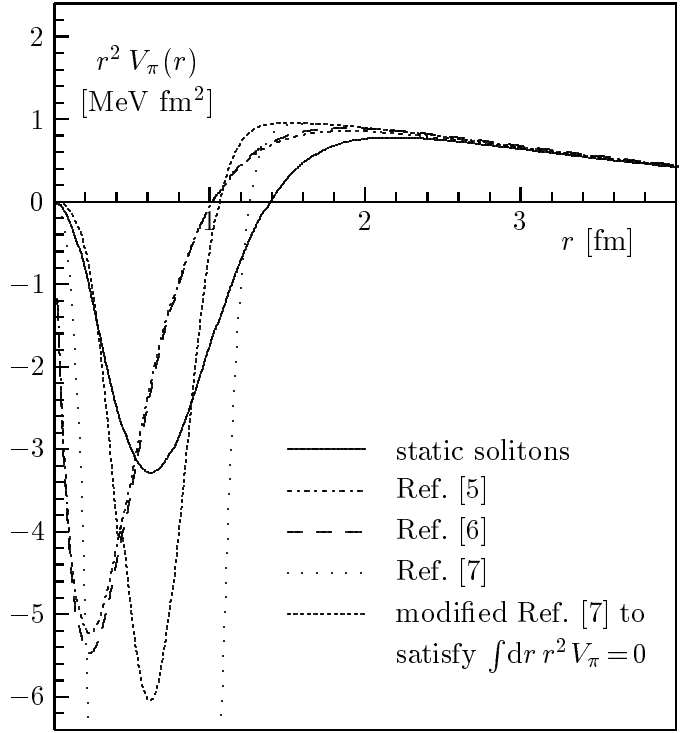}{90mm}
\end{center}
\caption{The pion exchange effective potential (multiplied by $r^2$)
between two quark solitons for $g=6$ (solid line) compared to some
phenomenological potentials.}
\end{figure}

Since our calculation gives just a crude picture of the potentials,
we have made only a simple estimate of the size of the nucleon and 
the $\Delta$(1232) by assuming harmonic oscillator wave functions 
$\varphi_{0s}(r)=(1/\pi b^2)^{3/4}\,\mathrm{e}^{-r^2/2b^2}$ 
to describe the orbital motion of each QS and determined 
the parameter $b$ variationally.
It is interesting to notice that both baryons are bound
even without including the confining potential.
Using $g=6$ the RMS radius of the orbital motion is 
$\sqrt{\langle{r^2_N}\rangle}=0.61$~fm for the nucleon and 
$\sqrt{\langle{r^2_\Delta}\rangle}=0.81$~fm for the $\Delta$;
the energy difference between the nucleon and the $\Delta$ is 
only 43~MeV.
Including a linear confining potential $V(r_{ij})=\kappa r_{ij}$
and using a popular value $\kappa = 0.5$~GeV/fm,
reduces the RMS radii to $\sqrt{\langle{r^2_N}\rangle}=0.36$~fm
and $\sqrt{\langle{r^2_\Delta}\rangle}=0.38$~fm; while the
energy difference increases to  125~MeV.
Using the values for the internal sizes of the U and D quark
the charge radii of the proton and the neutron are estimated 
and displayed in Table~1.

Let us stress that the strength of the attractive part of $V_\pi$
is almost entirely determined by the condition
$\int{\rm d}r\,r^2 V_\pi(r)=0$ which holds for any pseudoscalar
meson exchange effective interaction, and through this condition
to the value of the quark-pion coupling constant which is, 
in turn, constrained by the experimental value of $g_{\pi NN}$.
By shrinking the width of the attractive part (see e.g. Fig.~2),
its strength can be increased but not to the extent to
bring the $\Delta$-N splitting up to the experimental value.

\section{Conclusions}

We have shown that the linear $\sigma$-model is a suitable
framework for the microscopic description of constituent quarks.
It requires no new model parameters and gives sensible
predictions about the effective interaction between constituent 
quarks as well as about different static properties of the baryon.
While the present results are still based on a rather simplified
approach designed to explore the trends,
their relative success encourages the use of more
sophisticated techniques.

The pion-exchange effective potential between constituent 
quarks has a similar shape as the phenomenological potential 
used in Ref \cite{Riska,Paco,Fdelar,Glozman}.
Since a strong attractive part has been shown \cite{Glozman} 
to be needed to give simultaneously a good N-$\Delta$ splitting 
as well as the correct ordering of the Roper and negative parity 
states we anticipate that some important mechanisms are still 
to be included in order to model this type of interaction.
Possible candidates are: 
removal of center-of-mass motion of each constituent quark, 
nonadiabadic corrections, three-body forces between 
constituent quarks, inclusion of $\rho$ mesons in the model.

We get a rather strong $\sigma$-exchange potential.
Its role in the linear $\sigma$-model may be also to mimic a part 
of the confining potential at low energies
as well as to strongly reduce the energy of the system 
so that the ``effective constituent mass'' of quarks 
in hadrons is considerably less than the calculated 
mass for an isolated constituent quark.

\medskip
We are pleased to acknowledge stimulating and helpful discussions
with L. Ya. Glozman, W. Plessas, D. O. Riska as well as
with M. Banerjee and T. Cohen. 
We are particularly indebted to M. Fiolhais for his explanation 
of the method of linear momentum projection.


\begin{thebibliography}{99}

\bibitem{MG}
 A. Manohar and H. Georgi, Nucl. Phys. {B234} (1984) 189.
\bibitem{ChengLi}
 T. P. Cheng and Ling-Fong Li, Phys. Rev. Lett. {74} (1995) 2872.
\bibitem{Pirner}
 S. Baumgartner, H. J. Pirner, K. K\"onigsmann and B. Povh,
  Z. Phys. A 353 (1996) 397.
\bibitem{Riska}
 L. Ya. Glozman, D. O. Riska, Phys. Reports 286 (1996) 263.
\bibitem{Paco} 
 A. Valcarce, P. Gonz\'alez, F. Fern\'andez, V. Vento,
  Phys. Lett {B} 367 (1996) 35;
 D. R. Entem, A.I. Machavarini, A. Valcarce, A. J. Buchmann, 
  A. Faessler, F. Fern\'andez, Nucl. Phys. {A602} (1996) 308.
\bibitem{Fdelar} 
 Z. Dziembowski, M. Fabre de la Ripelle, and G. A. Miller,
  preprint nucl-th/9601022.
\bibitem{Glozman} 
 L. Ya. Glozman, Z. Papp, W. Plessas, Phys. Lett. {B} 381 (1996) 311.
\bibitem{Yoshi} 
 Y. Fujiwara, C. Nakamoto, and Y. Suzuki,   Phys. Rev. Lett. 76 
  (1996) 2242, KUNS-1395 (1996), Kyoto University.
\bibitem{Sigma} 
  M. Gell-Mann and M. L\'evy, Nuovo Cim. {\bf 16} (1960) 705.
\bibitem{BB} 
 M. C. Birse and M. K. Banerjee, Phys. Lett. B 
  {136} (1984) 284; Phys. Rev. D {31} (1985) 118.
\bibitem{KRS} 
 S. Kahana, G. Ripka and V. Soni, Nucl. Phys. A 
  {415} (1984) 351.
\bibitem{expsigma}
 N. A. T\"ornqvist and M. Roos, Phys. Rev. Lett. {76} (1996) 1575.
\bibitem{GR} 
 B. Golli and M. Rosina, Phys. Lett. B {165} (1985) 347.
\bibitem{Mike} 
 M. C. Birse, Phys. Rev. D {33} (1986) 1934.
\bibitem{CFGR} 
 M. \v{C}ibej, M. Fiolhais, B. Golli and M. Rosina,
  J. Phys. G. {18} (1992) 49.
\bibitem{colsys} 
 U. Asher et al., A. C. M. Trans. Math. Software {7} (1981) 209.
\bibitem{AFGR}
 L. Amoreira, M. Fiolhais, B. Golli and M. Rosina,
  J. Phys. G. {21} (1995) 1657.
\bibitem{Vikram}
 V. Soni, Orsay preprint IPNO/TH86-85, unpublished.
\bibitem{Wilets}
 E. G. L\"ubeck, M. C. Birse, E. M. Henley, and L. Wilets,
  Phys. Rev. D {33} (1986) 234.
\bibitem{Neuber}
  T. Neuber and K. Goeke, Phys. Lett. B {281} (1992) 202.

\end{thebibliography}
\end{document}